# Quantifying real-world energy use and CO2 emissions of electric vehicles via a city-scale bottom-up framework

Shuhan Ge [1], Yanqiao Deng [2], Minda Ma [3 *]

1. School of Automation and Intelligent Sensing, Shanghai Jiao Tong University, Shanghai, 200240, P. R. China
2. School of Future Technology, South China University of Technology, Panyu District, Guangzhou 511442, China
3. School of Architecture and Urban Planning, Chongqing University, Chongqing, 400045, P. R. China

Corresponding author: Prof. Dr. Minda Ma, Email: minda.ma@cqu.edu.cn

Homepage: https://globe2060.org/MindaMa/

## Graphical Abstract

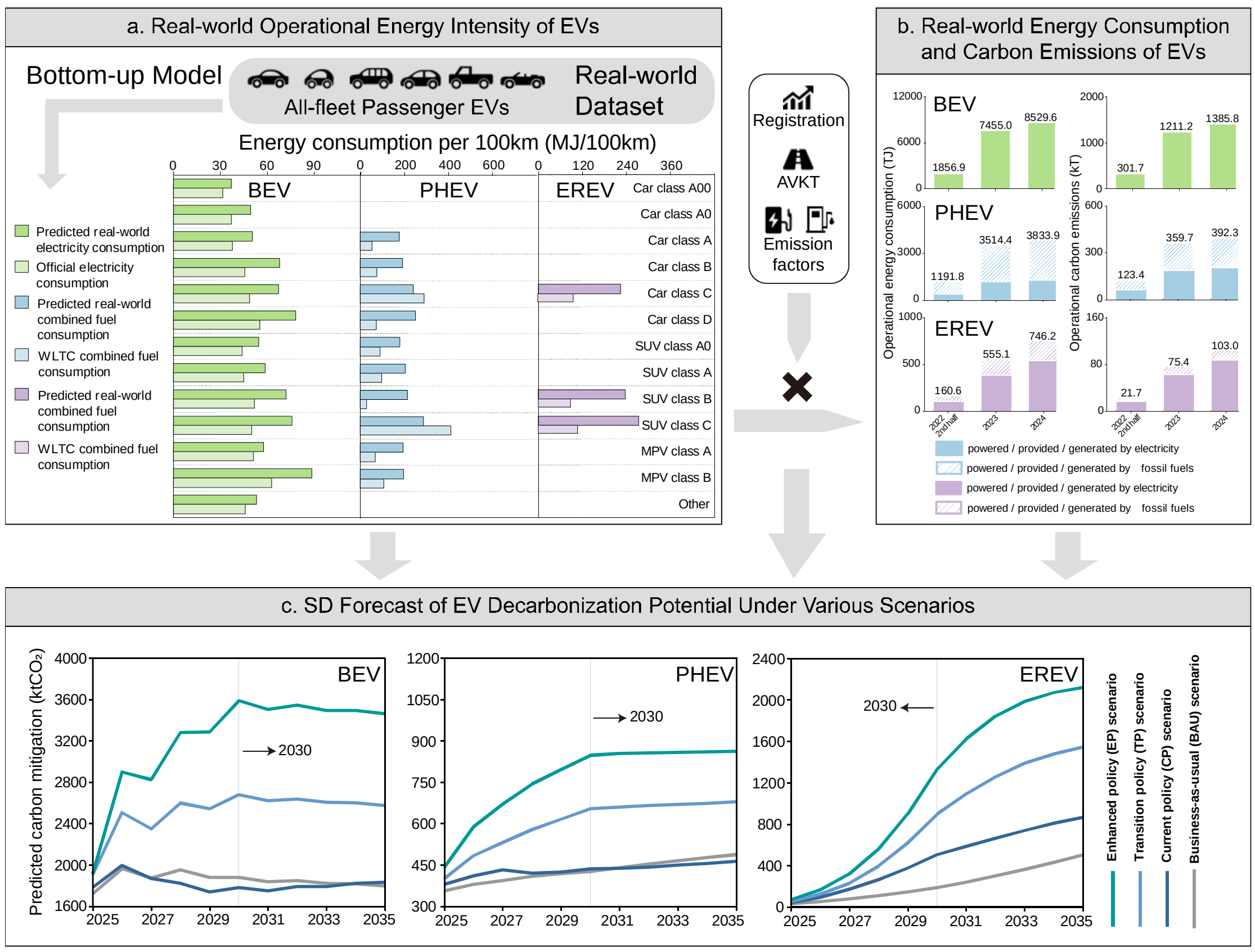

**Graphical abstract.** (a) Methods and materials of the proposed bottom-up energy intensity of all-fleet passenger EVs; (b) real-world energy consumption and carbon emissions of EVs; (c) system dynamic outlook model of the decarbonization potential of EVs in the coming decade.

## Highlights

- A bottom-up model is developed to estimate the real-world energy use and carbon emissions of EVs in Shanghai.
- On average, the real-world energy consumption of EVs is 36.2% greater than that of the official test data.
- Grid emission is the primary source of operational EV $CO_2$ emissions, generating up to 975 kt of $CO_2$ in 2024.
- Charging convenience, congestion and grid cleanliness are decisive for maximizing EV decarbonization.
- An outlook model for stakeholders to plan effective and sustainable decarbonization strategies has been developed.

## Abstract

Although electric vehicles (EVs) are scaling rapidly, city-scale evidence on real-world operational energy use and carbon dioxide ($CO_2$) emissions from EVs remains limited. Using Shanghai as a case study, this study develops a bottom-up framework covering all EV models registered between July 2022 and December 2024 to quantify model-specific real-world energy intensity, the operational energy mix, and associated $CO_2$ emissions. The results indicate that (1) pronounced and systematic underestimation by test-cycle values: on average, real-world use is 20.8% greater for battery electric vehicles (BEVs) and ~55% greater for plug-in hybrid electric vehicles (PHEVs), whereas extended-range EVs (EREVs) show the largest gaps, as many models consume 3.75 times more energy than their official data suggest. (2) From 2022–2024, electricity supplies more than 70% of operational energy, and power-sector emissions dominate EV operational $CO_2$, contributing 75.3%, 85.7% and 87.0% in 2022, 2023 and 2024, respectively. (3) BEVs achieve the greatest absolute mitigation under current policies, with 1,834 kilotons (kt) of $CO_2$ in 2035, modest benefits from PHEVs, and strong gains for EREVs under more ambitious policies (up to 2,122 kt of $CO_2$ in 2035). These findings underscore the need to align fleet electrification with grid decarbonization, alleviate congestion, improve charging accessibility, and narrow test-cycle versus on-road performance gaps to fully realize the climate benefits of EVs in megacities.

## Abbreviation notation

AER – All-electric range

AVKT – Annual vehicle kilometers traveled

BC – Battery capacity

BEV – Battery energy vehicle

CD – Charge-depleting

CLTC – China light-duty vehicle test cycle

CS – Charge-sustaining

EV – Electric vehicle

EREV – Extended-range electric vehicle

$ktCO_2$ – Kilotons of carbon dioxide

LCA – Life cycle analysis

MLR – Multilinear regression

MJ – Megajoules

$MtCO_2$ – Megatons of carbon dioxide

NEDC – New European driving cycle

PHEV – Plug-in hybrid electric vehicle

SD – system dynamic

SOC – State of charge

UF – Utility factor

WLTC –Worldwide harmonized light vehicles test cycle

## Nomenclature

$AER_{WLTC,j}$ – All-electric range of vehicle model $j$ under WLTC standard (unit: 100 km)

$AVKT_j$ – Annual electric vehicle kilometers traveled of vehicle model $j$ (unit: 100 km)

$Battery\ Capacity_j$ – Battery capacity of the vehicle model $i$

$EC_{i,j}$ – Total AVKT-based energy consumption of vehicle model $j$ with powertrain type $i$

$EI_{i,j}$ – Energy intensity of vehicle model $j$ with powertrain type $i$ (unit: kWh/100 km)

$EI_{CD,i,j}$ – Energy intensity of vehicle model $j$ with powertrain type $i$ under charge-depleting mode (unit: kWh/100 km)

$EI_{CS,i,j}$ – Energy intensity of vehicle model $j$ with powertrain type $i$ under charge-sustaining mode (unit: kWh/100 km)

$f_e$ – Carbon emission factor of electricity

$f_g$ – Carbon emission factor of gasoline

$Sal_j$ – Sales of vehicle model $j$

$UF_{CD}$ – Utility factor

$\zeta_e$, $\zeta_g$ – the electricity and gasoline conversion factor

$\xi_e$ – the unit conversion factor

# 1. Introduction

## *1.1. Background*

Our traffic system is experiencing great electrification, with 17 million electric vehicles (EVs) sold worldwide in 2024, 20% more than in 2023 [1]. This trend has been even more significant and dominant in China, the leading country in terms of the production, sales, and ownership of EVs, since 2015 [2]. In 2024, the market share of EVs could reach 45% in China [3], which offers considerable theoretical potential for reducing carbon dioxide ($CO_2$) emissions if electricity mixes are green enough [4, 5]. However, although EVs are universally considered eco-friendly and are believed to contribute significantly to the mitigation of greenhouse gas (GHG) emissions, several studies have revealed that electrification might not instantly lead to effective decarbonization [6, 7], especially the underestimation of EV emissions during the operational stage [8]. As a metropolis with the largest EV fleet worldwide, Shanghai offers a complicated yet comprehensive background for city-level estimation.

## *1.2. Literature review*

The estimation and prediction of energy consumption and carbon dioxide emissions for EVs have garnered significant scholarly interest, and a multitude of methodologies have been employed to address this complex issue in recent years, including vehicle dynamic simulations [9], computational intelligence (including data-driven prediction models [10] and machine learning approaches [11]), theoretical models [12] and statistical models [13].

First, vehicle dynamic simulation is among the most influential approaches in the early stages of EV energy and carbon estimation research. This method, which uses simulation platforms such as Autonomie [14, 15] and Greet [16], incorporates multiple factors, including ambient temperature [17], road conditions [18], and driving behaviors [19, 20], into its energy consumption predictions. However, as the simulation process tends to be specific to certain tested models, studies are often limited to a single type of vehicle [21]. Although it is theoretically possible to expand simulation results into a comprehensive, large-scale model, doing so requires astronomical amounts of high-quality data [15, 22].

Similarly, theoretical models, typically using life cycle assessment (LCA) methods [23], can estimate EV energy consumption and carbon emissions from a technical perspective. Through Cradle-to-Grave analysis [5, 24], these models encompass the entire lifecycle of EVs, including manufacturing, operation, and recycling, providing a comprehensive estimation. Additionally, these models can be applied at the country or even global level [25, 26], highlighting regional differences [27, 28]. However, during the operational stage, theoretical models often rely on simplified assumptions, leading to potential limitations in their accuracy [29].

In recent years, computational intelligence methods have attracted significant attention in the fields of EV energy consumption and carbon emission estimation. These methods are increasingly favored for their ability to handle complex, nonlinear relationships between variables and for their high accuracy in real-time predictions [30, 31]. These methods offer high accuracy without the need for detailed physical models, making them suitable for large-scale applications [32]. However, challenges remain, including the reliance on large datasets or exclusively measured real-world data for training [33, 34], which may limit generalization, especially under novel conditions.

Statistical models can generally be classified into two categories: top-down models and bottom-up models [35, 36]. The top-down approach focuses on macrolevel data and trends and uses national or regional statistics to estimate the impact of EV adoption on energy demand and carbon emissions [37]. They typically employ aggregate data such as national sales figures, economic growth, and energy consumption patterns to forecast broader trends [38, 39]. However, they often lack the granularity needed to account for local variations and may be subject to the biases inherent in large-scale statistical data [40]. On the other hand, the bottom-up method involves more pellets, starting at the level of detailed vehicle-specific data, such as real-world energy consumption and local driving behavior [41]. Therefore, they offer more precise insights into energy usage and carbon emissions under various conditions.

### *1.3. Motivation, contributions, and framework*

On the basis of the above review, two significant research gaps hinder accurate and comprehensive estimations of EV energy consumption and decarbonization performance. First, previous studies have largely relied on data from a single-vehicle model or a small group of models, which leads to substantial errors when applied to an entire fleet at the city level [42, 43]. Moreover, current models either overlook detailed travel characteristics or require locally customized datasets, severely limiting their scalability [44, 45]. Given the rapid pace of vehicle electrification, an approach that comprehensively covers existing EV models to accurately assess their operational efficiency and true environmental impact is needed [46]. Three key issues emerge from these gaps, which warrant further investigation:

- How can real-world energy use and carbon emission models be established for all EVs?
- What are the discrepancies in energy use between real-world EV operations and official data?
- How can the decarbonization potential of all EVs be accurately forecasted over the next decade?

This study aims to address these issues by providing a detailed analysis of the energy use and carbon emissions of EVs from 2022–2024, along with a decarbonization performance forecast from 2025–2035. This study is conducted at the city level, with Shanghai chosen as a case study because of its high global EV penetration rate. The primary motivation behind this research is to develop a comprehensive, transferable model for assessing the real-world performance of BEVs, PHEVs, and EREVs in terms of energy consumption and emissions. This is crucial for policymakers seeking to optimize strategies that maximize the environmental benefits of these vehicles. By analyzing the operational energy intensity and associated carbon emissions of various EV models, this study contributes to a more nuanced understanding of how these vehicles perform in practice, particularly in a dynamic city such as Shanghai, where both the electricity grid and infrastructure are rapidly evolving.

**The principal contributions of this study are threefold.** First, this work develops a bottom-up model that estimates real-world energy intensity and carbon emissions for different types of EVs in Shanghai, utilizing a dataset based on owner-measured data and registration statistics. Compared with official test cycle data, this model allows for a more accurate understanding of operational energy consumption, which tends to overestimate EV efficiency. In addition, the research introduces a detailed prediction model for future decarbonization performance under a series of scenario settings in Shanghai, offering valuable insights for future policy and infrastructure development.

The structure of this paper is as follows: Section 2 outlines the methods and materials, focusing on the multilinear regression model used for estimating real-world energy intensity, followed by the calculation of carbon emissions for each vehicle category. Section 3 presents the results and analyzes the energy consumption, carbon emissions, and sales trends of EVs in Shanghai. Finally, Section 4 discusses the implications of these findings and offers policy recommendations for improving the efficiency and decarbonization potential of EVs in the region.

## 2. Methods and materials

A bottom-up model was developed in an attempt to estimate the energy intensity and emissions of EVs in Shanghai during the research period, which lasted from July 2022 to December 2024. Section 2.1 elaborates on the multilinear regression (MLR) model developed from existing real-world owner-measured datasets. Section 2.2 introduces the estimation of the carbon emission intensity of EVs in operation. Section 2.3 introduces the datasets used in this study.

### *2.1. Estimation of real-world energy intensity*

Given that energy intensity is, in most cases, strongly related to several factors [47, 48], a multilinear regression model is applied in our research to estimate the real-world energy intensity of EV models as follows:

$$\hat{y} = \sigma_0 + \boldsymbol{\sigma} \cdot \boldsymbol{X} \tag{1}$$

where $\hat{y}$ denotes the predicted real-world electricity (for BEVs, in kWh/100 km) or fuel consumption (for PHEVs and EREVs, in L/100 km). $\boldsymbol{X}$ refers to selected independent variables that are subsequently determined through Pearson correlation analysis. $\boldsymbol{\sigma}$ is the coefficient vector trained on the dataset, whereas $\sigma_0$ is the intercept term accounting for the baseline offset.

For each EV category (BEV, PHEV, or EREV), this study first applies Pearson correlation analysis to the owner-measured dataset to evaluate the linear relationship between 10 candidate independent variables and real-world energy intensity. Five variables with relatively strong correlation coefficients $X = \{X_1, X_2, X_3, X_4, X_5\}$ were selected as predictors for the subsequent regression model, ensuring that the multilinear regression (MLR) models were built upon statistically relevant input features and enhancing model interpretability and performance.

If needed, the missing values among the selected $\boldsymbol{X}$ values were imputed via linear interpolation prior to model fitting. Furthermore, in an attempt to convert electricity consumption and combined fuel consumption into energy intensity with a unified unit, different conversion terms were added to the equation:

$$EI_{BEV} = \zeta_e \cdot (\sigma_{0,BEV} + \boldsymbol{\sigma_{BEV}} \cdot \boldsymbol{X_{BEV}}) \quad (2)$$

$$EI_{PHEV} = \zeta_g \cdot (\sigma_{0,PHEV} + \boldsymbol{\sigma_{PHEV}} \cdot \boldsymbol{X_{PHEV}}) \quad (3)$$

$$EI_{EREV} = \zeta_g \cdot (\sigma_{0,EREV} + \boldsymbol{\sigma_{EREV}} \cdot \boldsymbol{X_{EREV}}) \quad (4)$$

where $EI_i$ represents the predicted real-world energy intensity (unit: MJ/100 km) for models of powertrain $i$ (BEV, PHEV and EREV), $\zeta_e$ represents a unit conversion term that turns kWh/100 km to MJ/100 km, and $\zeta_g$ represents a unit conversion term that turns L/100 km to MJ/100 km. $\boldsymbol{\sigma_i}$ is the coefficient vector of powertrain $i$ trained on the dataset, whereas $\sigma_{0,\mathrm{i}}$ is the intercept term accounting for the baseline offset of powertrain $i$.

*2.2. Estimation of real-world energy consumption and carbon emissions*

The process of energy consumption can be fully described as follows. For BEV models, electricity is the only power involved during the operational phase. Therefore, on the basis of the estimation of energy intensity through the MLR model introduced in Section 2.1, their energy consumption can be obtained as follows:

$$EC_{BEV,j} = EI_{BEV,j} \times Sal_j \times AVKT_j \times 0.01 \quad (5)$$

where $EC_{BEV,j}$ refers to the newly added energy consumption for the BEV car model $j$ (unit: MJ) and where $EI_{BEV,j}$ represents its corresponding predicted real-world energy intensity (unit: MJ/100 km). $Sal_j$ indicates the annual sales of model $j$, which is derived from mandatory transportation insurance data released by the government. Finally, $AVKT_j$ quantifies the annual distance traveled by the vehicle in model $j$ (unit: km).

For the PHEV and EREV models, both electricity and fossil fuel contribute to propulsion during the operational phase. Their energy use can be conceptually divided into two distinct modes: the charge-depleting (CD) mode, during which the vehicle relies exclusively on electric power, and the charge-sustaining (CS) mode [49, 50], in which the internal combustion engine operates alongside the electric motor, maintaining the state-of-charge of the battery. Assuming that the proportion of the driving distance in the CD mode is denoted by $UF_{CD}$ (the proportion of the driving distance in the CS mode is $1 - UF_{CD}$), the real-world energy intensity for the PHEV/EREV model $j$ with powertrain type $i$ can be estimated as follows:

$$EI_{i,j} = EI_{CD,i,j} + EI_{CS,i,j} \tag{6}$$

where $EI_{i,j}$ denotes the predicted real-world energy intensity (unit: MJ/100 km) from the previous MLR model. $EI_{CD,i,j}$ and $EI_{CS,i,j}$ represent the estimated energy intensities under the CD and CS modes, respectively (unit: MJ/100 km). Additionally, $EI_{CD,i,j}$ can be calculated as follows:

$$EI_{CD,i,j} = \xi_e \times \frac{BC_j}{AER_{WLTC,j}} \times 100 \times UF_{CD} \tag{7}$$

where $BC_j$ and $AER_{WLTC,j}$ are official data obtained from Autohome (https://www.autohome.com.cn/), which refer to the battery capacity of model $j$ and the WLTC all-electric range of model $j$, respectively. $\xi_e$ serves as a unit conversion term that transfers kWh to the MJ. $UF_{CD}$, as mentioned, represents the proportion of the driving distance in the CD mode.

With $EI_{i,j}$ and $EI_{CD,i,j}$ already known, $EI_{CS,i,j}$ can be described as follows:

$$EI_{CS,i,j} = EI_{i,j} - EI_{CD,i,j} \tag{8}$$

Furthermore, the final newly added energy consumption (unit: MJ) for car model $j$ with power type $i$ can be calculated as follows:

$$EC_{i,j} = \left[EI_{CD,i,j} \times UF_{CD} + EI_{CS,i,j} \times (1 - UF_{CD})\right] \times Sal_{i,j} \times AVKT_{i,j} \times 0.01 \tag{9}$$

For the estimation of $UF_{CD}$, a standard chart is given, in which $UF_{CD}$ is measured in terms of the vehicle battery capacity.

Similarly, the estimation of carbon emissions can be divided into two parts. Although BEV models, which run fully on electricity, do not generate carbon emissions themselves on the road, they still account for carbon emissions originating from electricity generation [51]. Therefore, their newly added carbon emissions for the BEV model $j$ ($CE_{BEV,j}$) (unit: $kgCO_2$) can be estimated as follows:

$$CE_{BEV,j} = f_e \times \frac{1}{\xi_e} \times EI_{BEV,j} \times Sal_j \times AVKT_j \times 0.01 \tag{10}$$

where $EI_{BEV,j}$ indicates the predicted real-world energy intensity (unit: MJ/100 km). $f_e$ refers to the grid emission factor in Shanghai, which is obtained from governmental documents. $\xi_e$ serves as a unit conversion term that transfers kWh to the MJ.

As discussed above, for PHEVs and EREVs, their CD mode and CS mode energy intensities (unit: MJ/100 km) are estimated. For the PHEV and EREV models, the $CO_2$ emissions come from both electricity consumption in the CD phase and fuel combustion in the CS phase. Thus, the total emissions are calculated by separately accounting for both phases. The electricity-related emissions are the same as those in the BEV case, whereas the fuel-related emissions are derived from fuel consumption and its carbon content. Hence, we have:

$$CE_{CD,i,j} = f_e \times \frac{1}{\xi_e} \times EI_{CD,i,j} \times AVKT_{i,j} \times UF_{CD} \times 0.01 \quad (11)$$

$$CE_{CS,i,j} = f_g \times \frac{1}{\xi_g} \times EI_{CS,i,j} \times AVKT_{i,j} \times (1 - UF_{CD}) \times 0.01 \quad (12)$$

where $CE_{CD,i,j}$ and $CE_{CS,i,j}$ refer to newly added carbon emissions generated during the CD and CS phases, respectively, from model $j$ that runs on power type $i$ (unit: $kgCO_2$). $EI_{CD,i,j}$ and $EI_{CS,i,j}$ represent the energy intensity during the CD and CS phases, respectively (unit: MJ/100 km). $f_e$ is again the grid emission factor in Shanghai, whereas $f_g = 2.37\ kg/L$ is the fuel emission factor. $UF_{CD}$ represents the proportion of the driving distance in the CD mode. Finally, $\xi_e$ and $\xi_g$ are unit conversion terms.

### *2.3. Datasets*

Shanghai, a leading EV hub in China with a robust industrial foundation, favorable policies, and diverse urban characteristics, was chosen as the case area. In this study, all EV models registered in Shanghai from July 2022 to December 2024, including BEVs, PHEVs, and EREVs, were analyzed. To assess the real-world operational energy intensity and carbon emissions of EVs in Shanghai, a comprehensive dataset covering the period from July 2022 to December 2024 was compiled.

The registration data for the EV models were sourced primarily from the mandatory transportation insurance data released by the government. This dataset allowed for the analysis of overall EV sales trends across different periods and the market share composition by vehicle type, as presented in Section 3.1. It also provides the annual sales for each model. Key vehicle specifications, including battery capacity and all-electric range (AER) under WLTC conditions, were obtained from Autohome (https://www.autohome.com.cn/). These official data points were crucial for estimating real-world electricity intensity for the BEV, PHEV, and EREV models. To estimate real-world energy intensity, a multilinear regression model was applied, utilizing an owner-measured dataset to determine the relationships between various independent variables and energy consumption. This dataset provided the basis for the predicted real-world electricity consumption for BEVs and the combined fuel consumption for PHEVs and EREVs, as shown in Fig. 2a. The real-world owner-measured energy consumption data used for correlation with the curb weight (Fig. 2b) also stem from this comprehensive owner-measured dataset. Additionally, the annual number of vehicle kilometers traveled (AVKT) for each model in Shanghai was obtained from data reported in previous studies and estimations based on existing methodologies [52]. The grid emission factor for electricity in Shanghai was obtained from governmental documents, and the fuel emission factor was set at 2.37 kg/L. The unit conversion terms, $\zeta_e$ and $\zeta_g$, for converting energy consumption units were obtained from the U.S. Department of Energy.

## 3. Results

In this section, the results and findings associated with the sales, operational energy intensity, and carbon emissions of the EV models in various Shanghai are analyzed. Section 3.1 presents the sales of EV vehicles. Section 3.2 explores the energy intensity characteristics of EV models. Section 3.3 reveals overall and detailed trends in total energy consumption and carbon emissions during the study period.

### *3.1. Registrations of EV models*

Fig. 1 shows a cascade of features related to new registrations of EVs in Shanghai. An overview of the sales figures during the research period is provided in Fig. 1a. In 2023, 366,440 EVs were sold, which represented just over 50% of the total vehicle sales in the city. This substantial increase, however, experienced a downturn in 2024, with total sales dropping to 303,155 vehicles. The breakdown of sales figures reveals the dominance of BEVs in 2023, with BEVs accounting for 85.5% of all newly sold EVs. This represents a significant increase in market share compared with their penetration rate of only 58% in 2022. Conversely, PHEVs and EREVs experienced a marked decline in both sales and market share in 2023, which likely correlates with the increasing availability of BEVs that have become more competitive in terms of range and charging infrastructure. Nevertheless, 2024 witnessed their revival. This is especially true for EREVs. Their sales constantly increased, reaching the highest value of the same period during the research period, and their market share doubled in 2024.

Moreover, a seasonal pattern emerges in EV sales, with a noticeable peak during the months of November and December. This trend reflects a common pattern of increased sales toward the end of the year, possibly driven by year-end promotions, government incentives, or consumer behavior. In contrast, a sharp decrease in sales typically occurs in January, potentially because of seasonal factors such as the Chinese New Year holiday, which can disrupt purchasing patterns. Notably, the last two months of the year contribute more than 30% of the total annual sales, which is approximately equivalent to the combined sales in the first five months of the year.

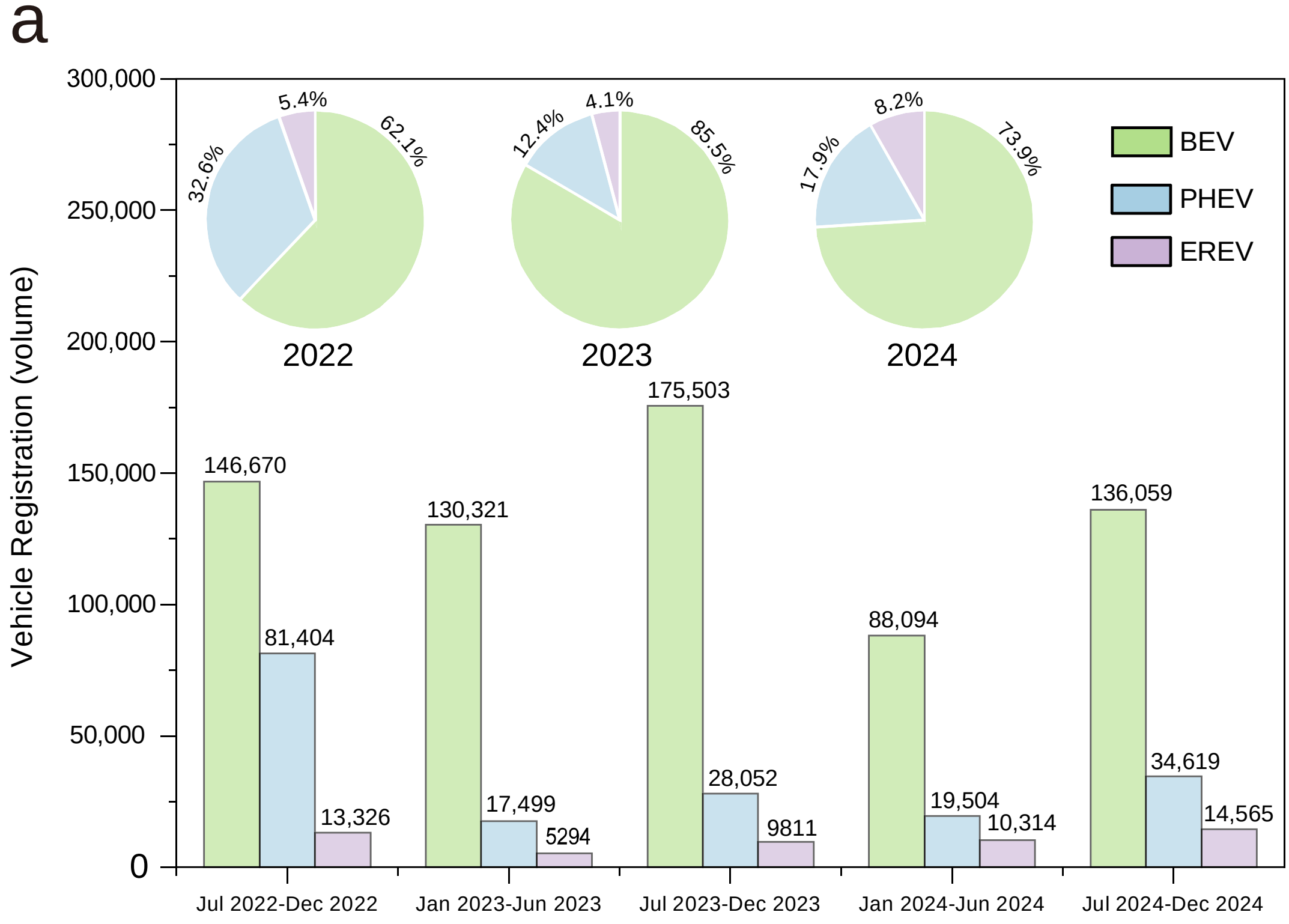

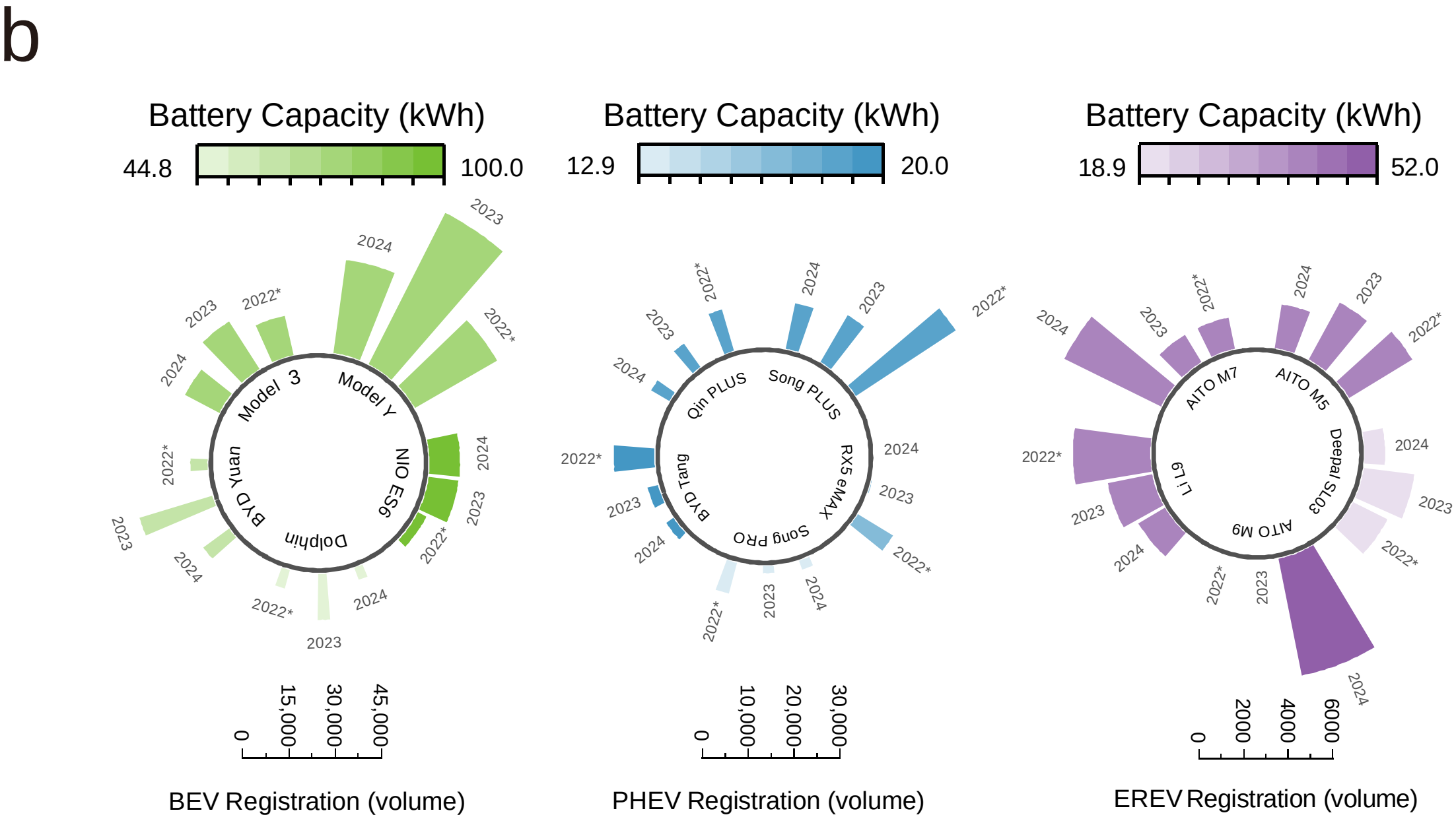

**Fig. 1** EV registrations in Shanghai for 2022 to 2024: (a) trends in the overall EV registration volume across different periods and market share composition by vehicle type; (b) model-level registrations analysis of the top-selling BEV, PHEV, and EREV models, including their battery capacities. Note: The year 2022 contains data only from July to December.

A model-level breakdown of sales trends, which contains a series of operational gasoline intensities of the top twenty selling models, is shown in Fig. 1b. It highlights the most popular BEV,

PHEV, and EREV models sold during the research period. The bar lengths, widths and color intensities offer insights into the sales volumes, retail prices and battery capacities of these vehicles. A positive correlation between battery capacity and retail price is illustrated in the chart, as expected. The analysis of PHEVs in Fig. 1b reveals that the most popular models in this category are typically economical vehicles, which are designed to offer a balance between performance and fuel efficiency and are affordable with typically moderate battery capacities that allow for short electric-only trips while still retaining the flexibility of a gasoline engine for longer journeys. In contrast, EREVs are more likely to be luxury or larger vehicles. EREVs, which combine the benefits of electric propulsion with a gasoline engine as a backup, are often marketed as premium options, featuring larger battery capacities to ensure longer electricity-only driving ranges. These vehicles are generally positioned at the higher end of the market, with substantial prices reflective of their advanced technology, larger size, and premium features. Finally, the BEVs in Fig. 1b present the greatest variety in terms of both battery capacity and retail price. The BEV segment encompasses a broad spectrum of vehicles, ranging from affordable compact cars to high-end luxury sedans and SUVs. This should be within anticipation, as BEVs cover the majority of EVs and enjoy a wide range of target customers.

*3.2. Energy intensity of operating EVs*

Fig. 2a shows a representative sample of EVs in operation between July 2022 and December 2024, covering the most popular models with different powertrain systems in each category, along with their official and estimated energy consumption data. Energy consumption is characterized by electricity consumption per 100 km for BEVs and combined fuel consumption per 100 km for PHEVs and EREVs. The estimated energy consumption varied significantly among the models, ranging from the most efficient Hongguang MINIEV at 37.20 MJ/100 km (MJ stands for mega joule) to the most energy-consuming Land Rover Defender at 281.75 MJ/100 km. Owing to their higher thermal efficiency, BEVs, which run completely on electric motors, have noticeably lower energy consumption than do PHEVs and EREVs, which rely partly on internal combustion engines. On average, BEVs achieve an estimated energy efficiency that is 161.51 MJ/100 km lower than that of PHEVs and 179.07 MJ/100 km lower than that of EREVs. When examined from a category perspective, even vehicles of the same size exhibit distinct differences in energy consumption. For example, according to the estimation, compared with their car counterparts with approximately the same curb weight, BEV SUVs consume 9.09% more energy. This figure increases to 23.1% for PHEVs and 15.5% for EREVs. Moreover, the estimated data indicate that MPVs are the most energy-demanding category, with BEV models requiring, on average, 14.0% more energy and PHEV models 29.6% more energy to cover 100 km than car benchmarks do. These results align with expectations for most models, as the taller bodywork of SUVs and MPVs contributes to lower aerodynamic efficiency and increased air drag, leading to excessive energy consumption.

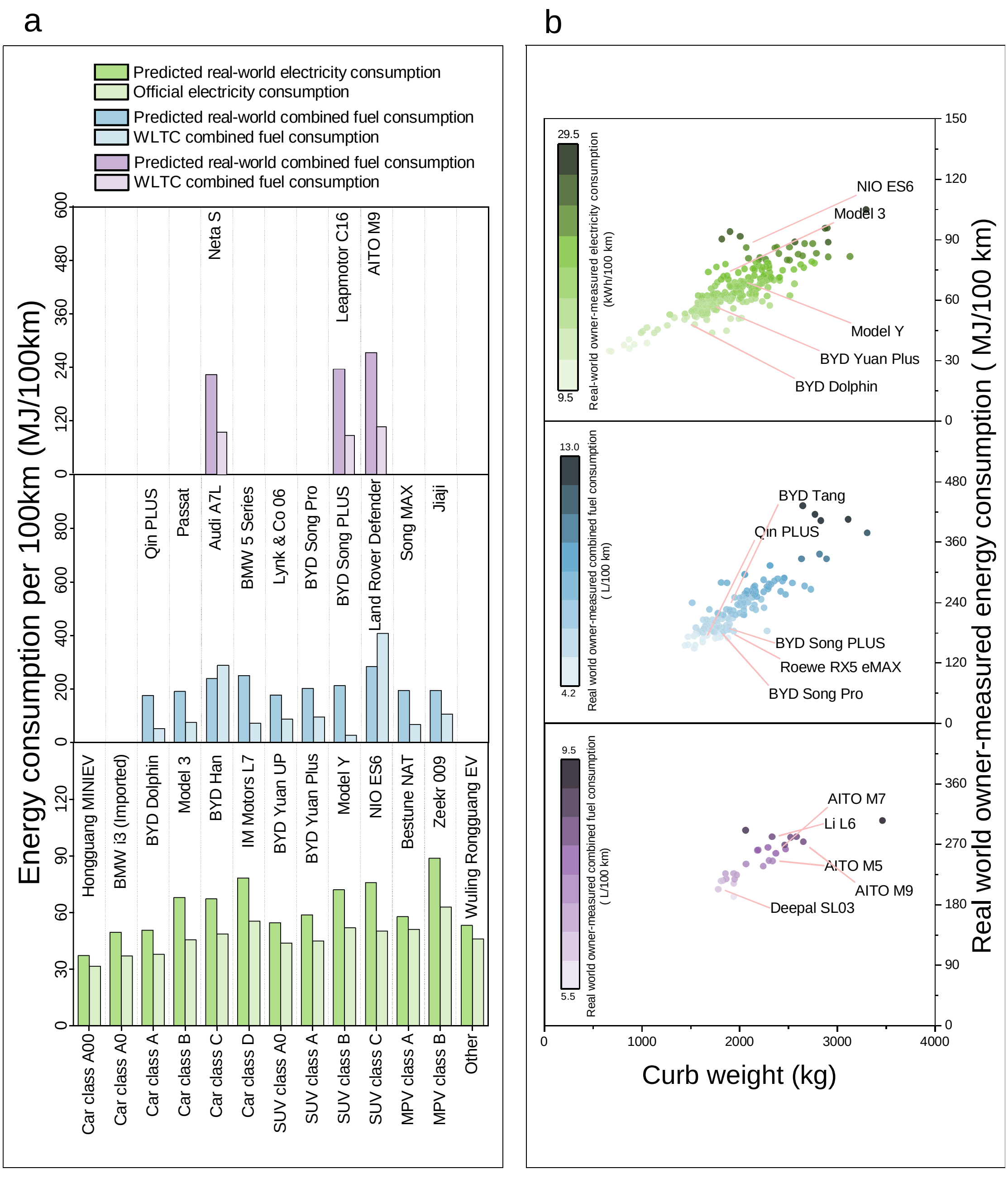

**Fig. 2** Energy consumption of EVs in Shanghai: (a) comparison of predicted real-world and WLTC energy consumption for various EV models, categorized by vehicle type; (b) correlation between curb weight and real-world owner-measured energy consumption for top-selling BEV, PHEV, and EREV models, alongside their respective battery/fuel consumption scales.

With respect to the differences between official energy consumption and predicted real-world data, a clear overestimation of efficiency among official data is observed. Overall, the predicted real-world owner-measured energy consumption is 36.2% higher than the official data released by the CLTC (for BEVs) or WLTC (for PHEVs and EREVs). Among these, the official data for EREVs appear to be the least reliable, as all 26 models consume, as the estimation suggests, on average, 1.38 times more combined fuel than claimed, with an average deviation rate of 87.4%. On the other hand, BEV models tend to have more accurate official data, which are 20.8% lower than the predicted real-world measurements. PHEVs, however, display the most scattered distribution. Although approximately 12% of the models have higher WLTC energy consumption than the predicted owner-measured values do, their energy efficiency is still generally overestimated by a margin of 55.2%. Moreover, SUVs and MPVs tend to exhibit more severe deviations, although not all the models exhibit this characteristic. Combining these findings, it is reasonable to assume a potential positive correlation between real-world energy consumption and the deviation rate.

Fig. 2b provides insight into the correlation between vehicles' real-world owner-measured energy consumption and their curb weight. In general, a strongly positive correlation is observed between these two features, as the Pearon correlation coefficient is set at 0.77, 0.85 and 0.73 for BEVs, PHEVs and EREVs, respectively. This finding fits with our previous knowledge that heavier vehicles consume more energy. However, the analysis also reveals that compared with heavier vehicles, lighter vehicles are more strongly correlated with curb weight and energy consumption. Specifically, vehicles with curb weights less than 2,000 kg have a Pearson correlation coefficient that is 0.08 greater than that of vehicles exceeding this threshold. This suggests that the relationship between weight and energy consumption is more consistent for lighter vehicles. For heavier vehicles, the correlation weakens, which may be explained by the greater diversity in additional equipment that these vehicles typically offer. Larger and heavier models often have advanced features, premium accessories, and larger battery capacities, all of which can introduce variability in energy consumption that is not directly related to curb weight, such as entertainment systems, larger wheels, and heavier battery packs.

In addition, vehicle type plays a role in this correlation. EREVs, which often have larger batteries and more complex drive train systems than BEVs or PHEVs do, may exhibit greater energy consumption variability. Similarly, PHEVs, which can switch between electric and gasoline power, could show an even wider range of energy consumption on the basis of how frequently the gasoline engine is used.

### *3.3. Energy consumption and carbon emissions of operating EVs*

On the basis of the aggregated operational data for BEVs, PHEVs, and EREVs from 2022–2024, clear temporal and intercategory differences in both energy use and carbon emissions can be observed in Fig. 3. Total operational energy consumption for all three categories has risen steadily over the three-year period, driven largely by increasing stock. The absolute increase in BEVs is greatest, with annual energy use increasing from approximately 1856.9 TJ in the second half of 2022 to 8529.6 TJ in 2024, representing a growth factor of more than 400%. This surge can be attributed not only to higher market penetration but also to consistently high AVKT values, which remained above 14,000 km per year on average. PHEVs and EREVs also exhibited year-to-year growth in total energy use, but their trajectories were less steep than those of BEVs, reflecting their smaller market shares and, in the case of EREVs, more limited deployment in the passenger vehicle sector.

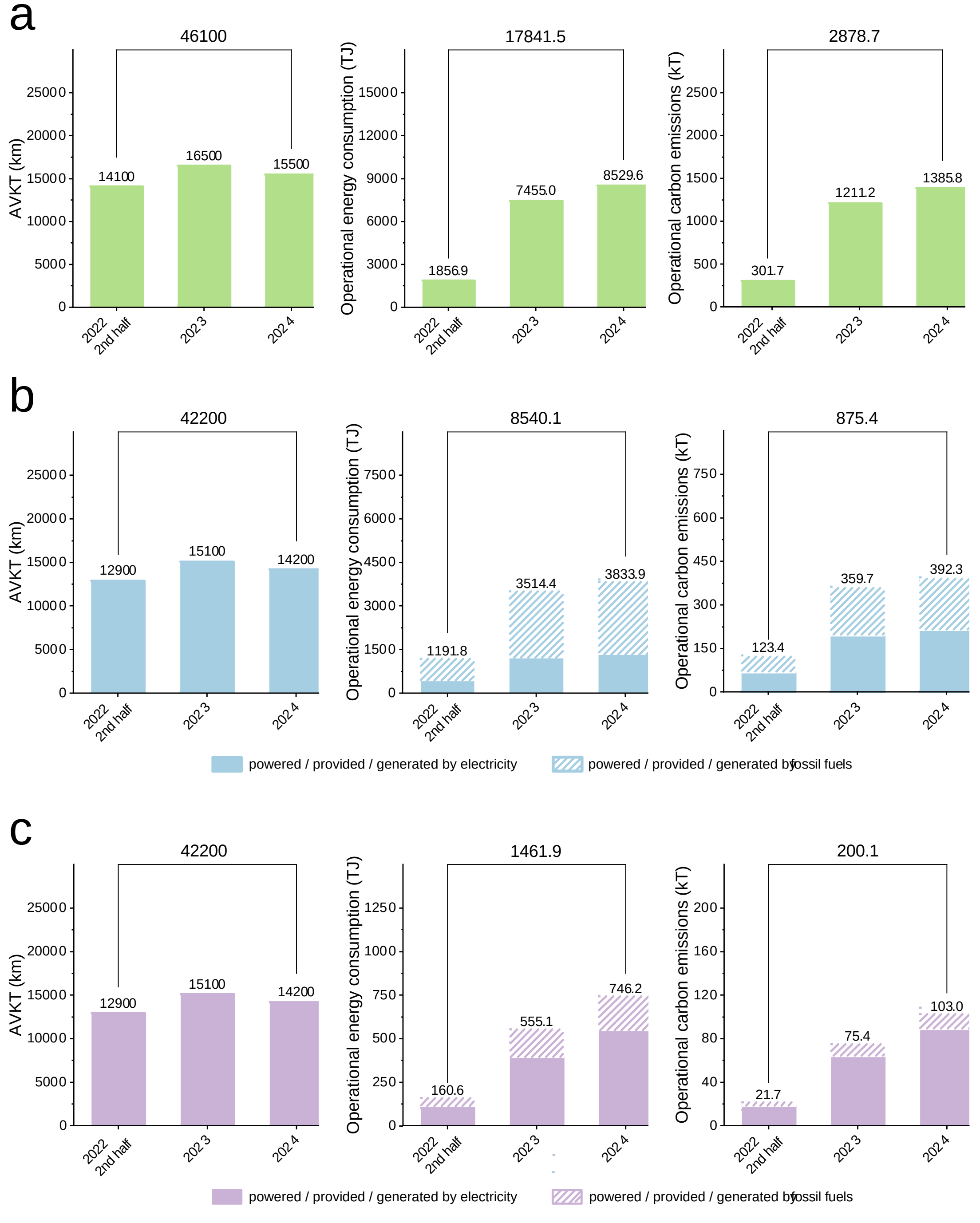

**Fig. 3** Annual vehicle kilometers traveled, operational energy consumption, and operational carbon emissions for (a) BEVs in Shanghai, (b) PHEVs in Shanghai, and (c) EREVs in Shanghai.

When the categories are compared, BEVs dominate the total energy demand despite being the only fully electric segment, which is a consequence of their larger fleet size and intensive usage patterns. In contrast, PHEVs, which operate in a blended mode between electricity and gasoline, consume less total energy but exhibit a higher share of fuel-derived energy, a characteristic that significantly shapes their carbon emission profile. EREVs, with their extended-range configuration and greater reliance on electric propulsion, maintain the lowest total energy consumption among the three categories and a significantly greater share of electric-powered energy than do PHEVs do.

From an emissions perspective, structural differences in energy sources create distinct carbon profiles. BEVs' carbon emissions originate entirely from electricity generation, making their climate impact directly dependent on regional power grid carbon intensity. In areas with less cleaner grids, such as Shanghai, which ranks only 19th in China, the gap narrows compared with fuel use categories. During the study period, total operational emissions from BEVs increased sharply in parallel with their energy use, reaching approximately 1385.8 kt$CO_2$ in 2024. PHEV emissions followed a similar but moderate trend, whereas EREV emissions remained relatively stable and low. These findings indicate that while BEV expansion is essential for long-term decarbonization, without concurrent improvements in grid cleanliness, their growing fleet size will continue to contribute significantly to total sectoral emissions. Moreover, targeted operational strategies for PHEVs and EREVs—such as increasing the electric driving share and improving hybrid system efficiency—could deliver near-term emission reductions, particularly in regions where clean electricity is already accessible.

The energy structure of EVs in Shanghai from 2022–2024, shown in Fig. 4, reveals a shift toward cleaner vehicle technologies. The proportions of energy consumption and carbon emissions from electricity and fossil fuels are shown in Fig. 4a, with a focus on AVKT, energy consumption, and carbon emissions. Throughout this period, electricity remains dominant, accounting for more than 70% of both energy consumption and carbon emissions. This dominance is driven primarily by the rise in BEVs, which operate entirely on electricity (utility factor = 1). However, in 2024, the market for EREVs, which combine electric motors and internal combustion engines, increased. This results in a slight increase in the proportion of AVKTs powered by fossil fuels. While fossil fuel usage has increased, it has not led to a proportional increase in energy consumption or carbon emissions.

Energy consumption and carbon emission analyses align with this trend. Although the share of electricity increases slightly because of the growing presence of EREVs, the emissions from electric vehicle operation remain relatively low. In contrast, energy consumption and emissions from fossil fuels—related to internal combustion engine use—decrease slightly as fossil fuel consumption decreases. This demonstrates that despite the increased fossil fuel component in AVKT, the adoption of electric vehicles and better grid efficiency continue to reduce overall carbon emissions, underscoring the role of electricity in lowering the environmental impact of transportation.

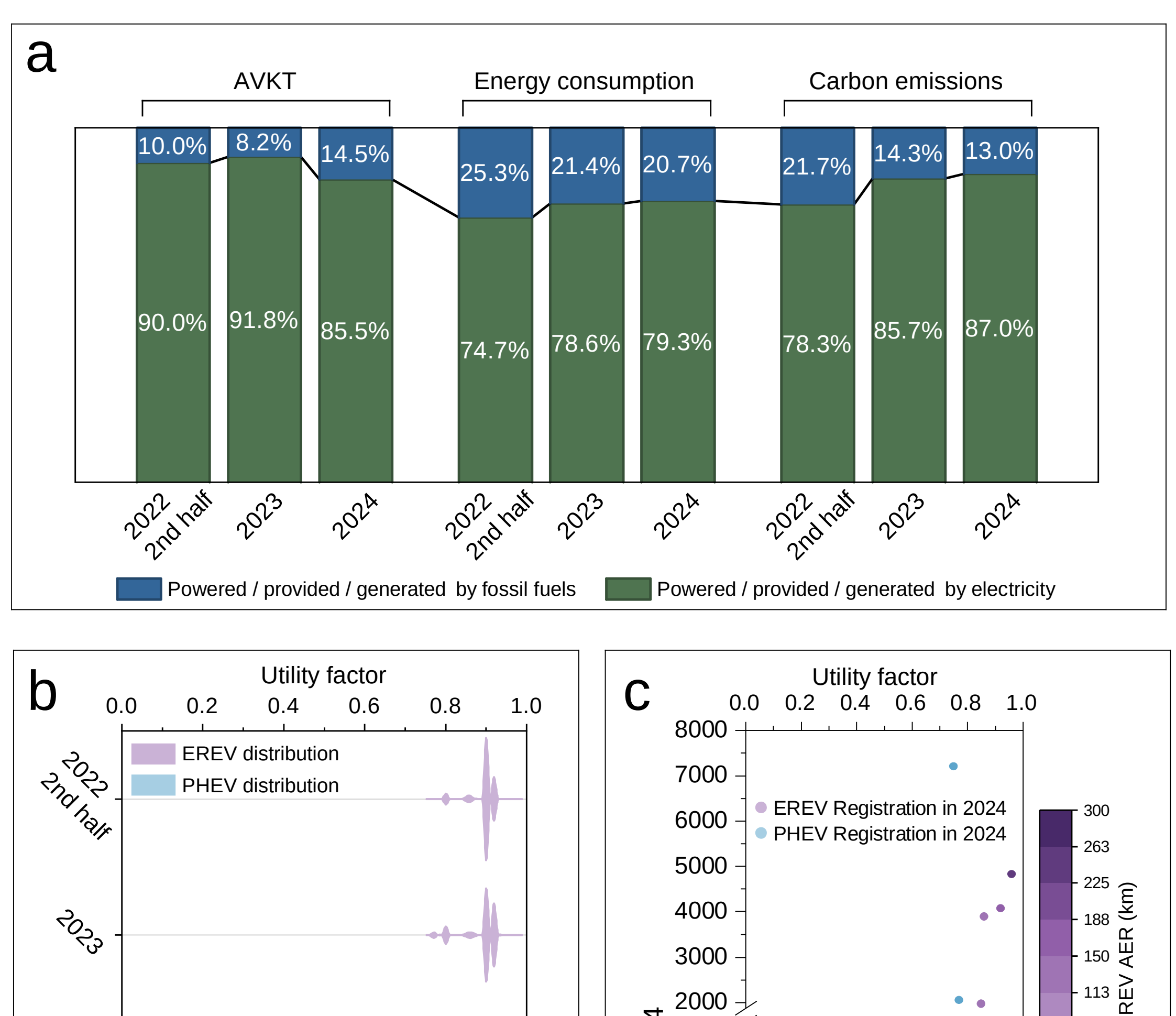

**Fig. 4** Power structure from the second half of 2022 to 2024: (a) proportions of AVKT, energy consumption, and carbon emissions powered by electricity and fossil fuels; (b) the distribution of EREVs

and PHEVs with respect to their utility factors from the second half of 2022 to 2024; (c) registration of EREVs and PHEVs in 2024 as a function of their utility factors, with color intensity reflecting the AER.

The utility factor (UF), which represents the proportion of driving time or distance a vehicle operates in electric-only mode, is a critical indicator of the effectiveness of EVs in reducing fossil fuel consumption. A violin plot comparing the UF distributions of EREVs and PHEVs from the second half of 2022 through 2024 is shown in Fig. 4b. The data reveal that compared with PHEVs, EREVs present a greater average utility factor, indicating a greater reliance on electricity for their operation. This is expected, as EREVs are typically designed with larger battery capacities and longer electric-only ranges than PHEVs, which can operate on both electricity and gasoline.

From 2022–2024, the utility factor for EREVs shows a relatively narrow distribution, with most models demonstrating high UFs, ranging from 0.77–0.99. On the other hand, PHEVs demonstrate a much wider range of UFs, reflecting more diverse usage patterns. Some PHEVs operate primarily on electricity, whereas others rely heavily on their internal combustion engine, resulting in lower utility factors. This greater variability in the utility factor for PHEVs highlights the diversity of user preferences and the adaptability of PHEVs to different driving conditions, including longer trips that necessitate gasoline usage.

Over the years, there has been an observable shift in the utility factor distribution for EREVs. The introduction of high-utility-factor models in 2024 marked a shift toward greener, more electricity-dependent vehicles, aligning with the broader market trend toward electrification. The PHEV fleet, however, has not experienced a similar shift, and in fact, a higher proportion of lower-utility factor models appears, suggesting that the trend in PHEVs may lean toward vehicles with a greater reliance on gasoline, possibly because of consumer preference for extended driving ranges. In Fig. 4c, the registrations of EREVs and PHEVs in Shanghai in 2024 are plotted against their utility factors, which could help explain the varied shifts among EREVs and PHEVs in 2024. The scatter plot visually demonstrates the relationship between the registration volume and the utility factor, with darker shades indicating higher AERs. These data underscore a pivotal trend: EREVs with higher AERs, indicating greater electric-only driving capabilities, are more frequently registered in 2024, whereas PHEVs with lower electric-only ranges are registered in larger numbers. The color gradient in the plot, which is associated with higher AERs with deeper colors, illustrates the market preference for vehicles with greater electric-only driving ranges, a preference that increasingly drives vehicle registrations.

The dominance of EREVs with high AERs reflects a broader shift in consumer behavior toward greener alternatives, driven by both government policies and growing awareness of environmental issues. On the other hand, the steady registration of lower-AER PHEVs suggests that a significant portion of the market still favors vehicles that combine electric propulsion with internal combustion engines for longer travel distances, particularly in regions where charging infrastructure may be less accessible.

## 4. Discussions

In this section, an outlook model based on the system dynamics approach is presented to evaluate the decarbonization performance of EVs in the Shanghai region for 2025–2035 under four distinct scenarios. The robustness of the model is subsequently examined, followed by a sensitivity analysis to assess its reliability. Finally, a set of policy implications is proposed on the basis of the insights derived from the model findings.

### *4.1. Carbon outlook*

As described in Section 2, the carbon emissions of EVs can be decomposed into four key factors: carbon intensity, vehicle stock, AVKT, and the power grid emission factor. Among these indicators, the carbon intensity of EVs and the projected cleanliness of the future power grid can be derived from governmental reports. Building upon existing studies [53, 54], a refined system dynamics (SD) model is then employed to predict the future stock of BEVs, PHEVs, and EREVs. Moreover, the AVKTs of these three categories of EVs for 2025–2035 are quantified via a newly developed SD model. By integrating these components, a comprehensive estimation model of EV carbon emissions is constructed. Finally, by subtracting the estimated emissions of EVs from those generated by an equivalent scale of conventional vehicles, the model enables an evaluation of the future decarbonization performance of EVs.

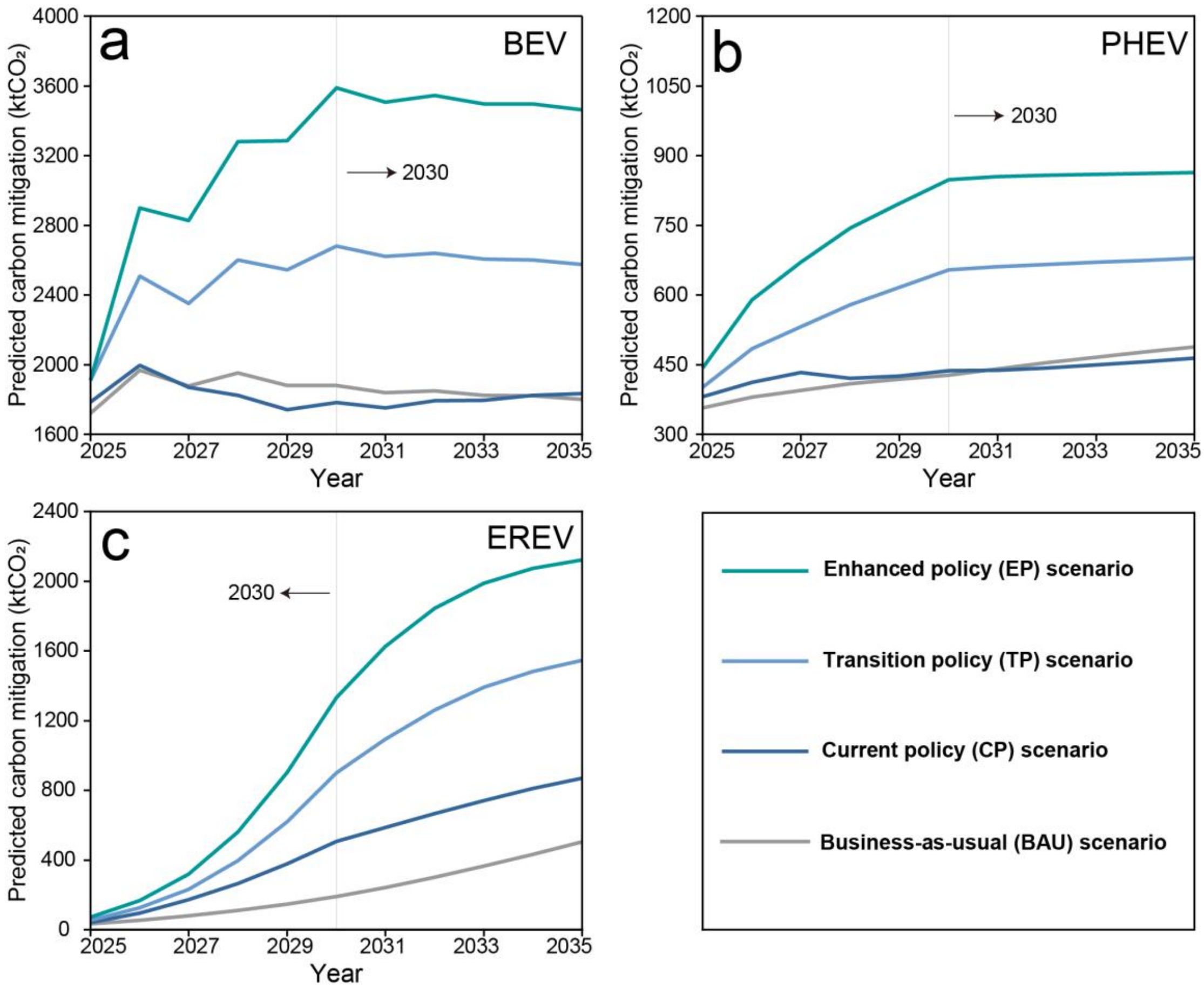

**Fig. 5** Predicted decarbonization performance of EVs in Shanghai for 2025 to 2035: (a) predicted decarbonization performance of BEVs; (b) predicted decarbonization performance of PHEVs; (c) predicted decarbonization performance of EREVs.

The projected carbon mitigation outcomes of EVs in the Shanghai region for 2025–2035 under four distinct scenarios are shown in Fig. 5: the baseline (BAU) scenario, the current policy (CP) scenario, which extends the same policy in operation to the year 2035; the transition policy (TP) scenario, which is set to achieve a carbon peak in 2035; and the enhanced policy (EP) scenario, which involves stronger policies than the TP scenario and aims at further decarbonization. Subfigures a, b, and c correspond to BEVs, PHEVs, and EREVs, respectively. Across all three vehicle categories, the results highlight both the magnitude of potential mitigation and the variation induced by different policy intensities.

For BEVs (Fig. 5a), the predicted mitigation effect makes the most substantial absolute contribution. Under the EP scenario, the mitigation potential rapidly increases, peaking at approximately 3,600 $ktCO_2$ by the early 2030s and maintaining a relatively stable level thereafter. In contrast, the BAU pathway results in only marginal improvements, stabilizing below 2,000 $ktCO_2$.

The CP scenario and TP scenario fall between these extremes, with the TP scenario consistently outperforming the CP scenario by approximately 300–500 kt$CO_2$. This pattern underscores the critical role of policy ambition in unlocking the decarbonization potential of BEVs, which already dominate the EV market in terms of both technology maturity and adoption rates.

In the case of PHEVs (Fig. 5b), the mitigation trajectory is more modest in scale. Even under the EP scenario, the cumulative mitigation levels plateau at approximately 900 kt$CO_2$, which are significantly lower than those of BEVs despite similar time horizons. The relatively constrained benefits can be attributed to the dual-fuel nature of PHEVs, where reliance on gasoline continues to limit carbon reduction. Nevertheless, the gap between scenarios is evident, with the mitigation achieved under the BAU scenario nearly doubling by 2035. These findings suggest that compared with BEVs, targeted policies can still enhance PHEV performance, although their long-term role in decarbonization appears to be secondary.

EREVs (Fig. 5c) display a distinctive growth pattern. Unlike those of BEVs and PHEVs, their mitigation potential increases steadily across the time horizon, reaching more than 2,000 kt$CO_2$ under the EP scenario by 2035. The continuous upward trajectory reflects the transitional nature of EREV technology, where gradual improvements in efficiency and grid decarbonization contribute to sustained emission reduction. Notably, the divergence between the BAU and EP scenarios widened significantly after 2030, highlighting the delayed but accelerating impact of stronger policy frameworks.

Taken together, the results of the comparative analysis reveal that while BEVs dominate in absolute mitigation, EREVs demonstrate the most dynamic growth under stringent policy scenarios, and PHEVs provide only limited supplementary contributions.

### *4.2. Robustness of the carbon outlook model*

In this study, an integrated system dynamics model was employed to estimate the future decarbonization performance of EVs, as outlined in Section 4.1. To assess the robustness of the model, the predicted vehicle stock and AVKT are compared with observed data derived from existing datasets and governmental reports. The comparison demonstrates that, under the CP scenario—which reflects the continuation of current policies—the model is strongly consistent with real-world data in terms of both vehicle stock and AVKT. This alignment enhances confidence in the accuracy of the subsequent decarbonization estimations. Furthermore, across different scenario settings, the model produces coherent and stable trends, with only minor fluctuations. Importantly, the relative hierarchy of the scenarios remains plausible, indicating that the model can reliably capture the expected variation in outcomes under differing policy assumptions.

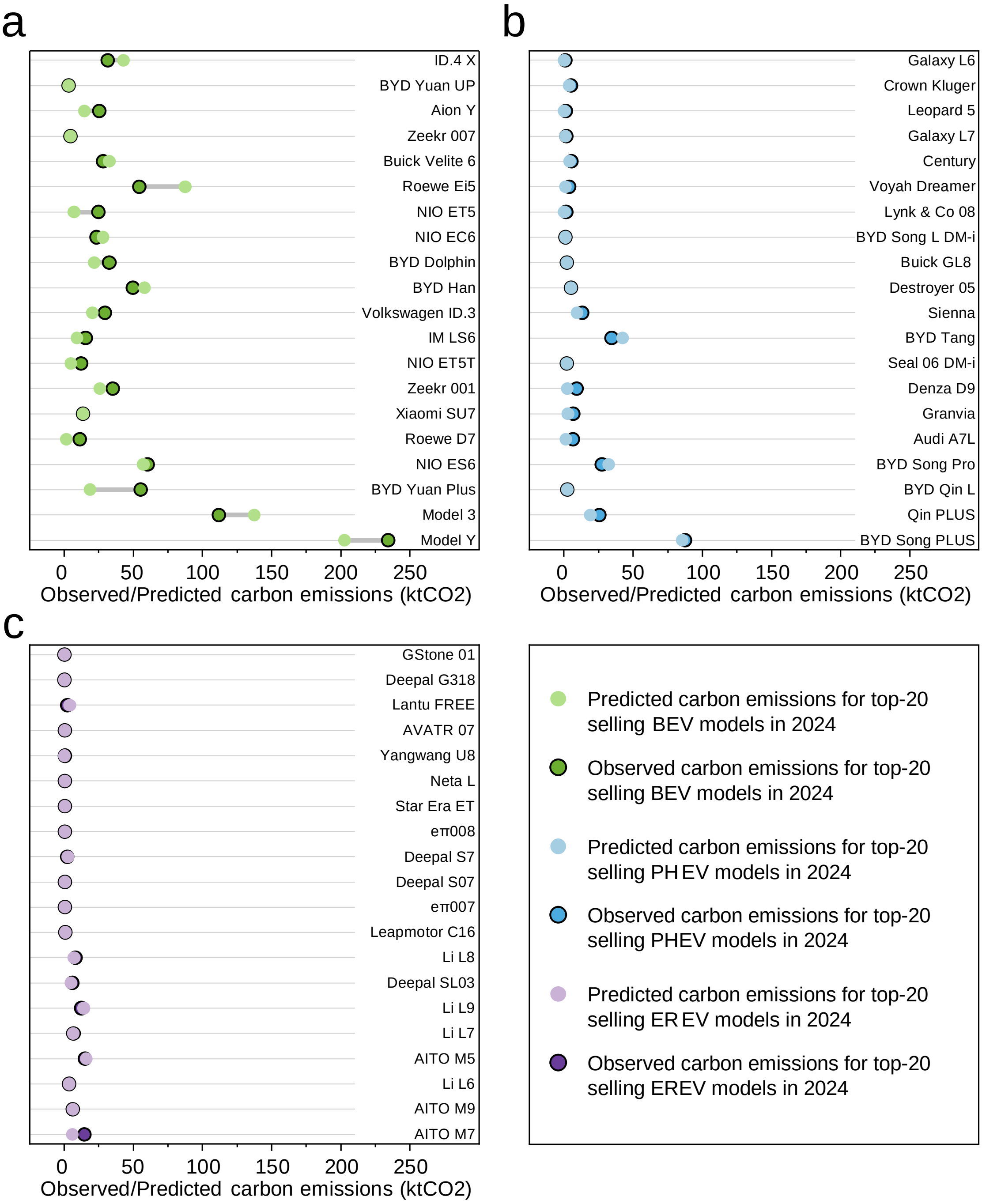

**Fig. 6** Observed and predicted carbon emissions of top-20 selling EVs in Shanghai in 2024: (a) observed and predicted carbon emissions of top-20 selling BEVs; (b) observed and predicted carbon emissions of top-20 selling PHEVs; (c) observed and predicted carbon emissions of top-20 selling EREVs.

More specifically, the AVKT estimation demonstrates strong consistency with real-world data, with a deviation of only –0.3%, whereas the stock estimation exhibits a slightly larger deviation of –5.9%. A closer inspection indicates that the majority of this discrepancy originates from the BEV stock. The relative underestimation can be explained by several factors. First, the structural assumptions embedded in the system dynamics model may lag in capturing the early-stage exponential growth of BEVs, thereby producing conservative projections [55]. Moreover, the model may not fully account for the accelerating influence of policy instruments such as purchase subsidies, license plate incentives, and infrastructure expansion, all of which drive faster-than-anticipated adoption [56]. Nevertheless, the gap between the model's estimates and the observed data narrows progressively over time, suggesting an improvement in long-term accuracy and reinforcing the model's reliability in predicting decarbonization performance through 2035.

### *4.3. Sensitivity analysis*

To further evaluate the robustness of the system dynamics model and disentangle the relative contributions of different external drivers to the carbon mitigation potential of EVs in Shanghai, a comprehensive sensitivity analysis was conducted. As illustrated in Fig. 7a, the tornado diagram provides a comparative overview of the extent to which variations in each parameter (set at 85% and 115% of their baseline values) influence the predicted mitigation outcomes. Four factors emerge as the most sensitive: the congestion index, charging convenience factor, power grid emission factor, and electric intensity, each exerting an impact magnitude between 6.8% and 9.1%. In contrast, parameters such as fuel cost, travel demand, and fuel consumption at the minimum battery state of charge (SOC) alter the mitigation results by less than 2%, suggesting their relatively limited importance within the current modeling framework.

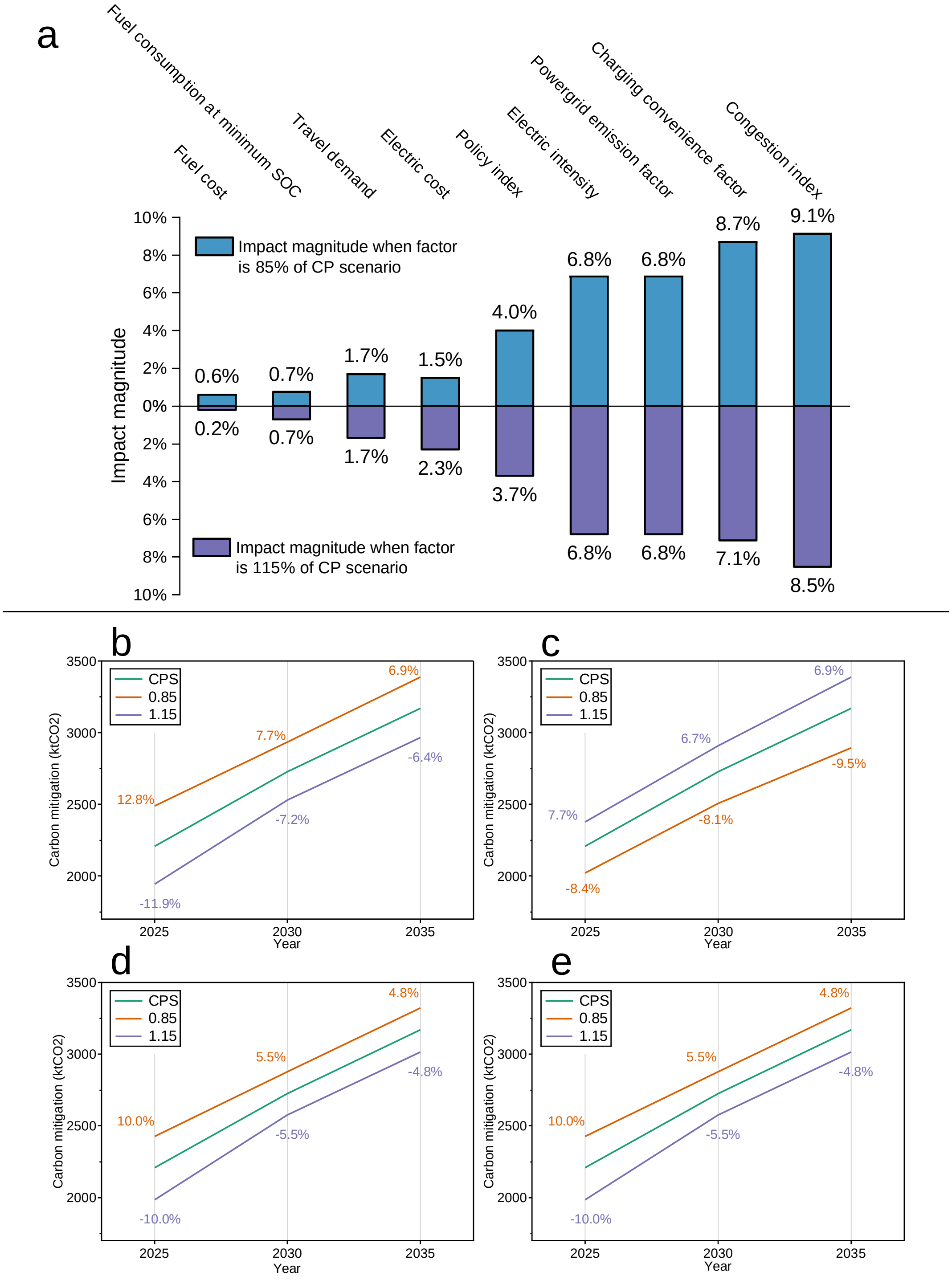

**Fig. 7** Sensitivity analysis of the carbon mitigation potential of EVs in Shanghai: (a) tornado diagram illustrating the relative influence of different external factors under 85% and 115% parameter variations; (b) effect of the congestion index on the projected carbon mitigation outcomes; (c) effect of the charging convenience factor; (d) effect of the power grid emission factor; (e) effect of electricity intensity.

A more detailed examination of the four most sensitive factors is presented in the following figures. The congestion index (Fig. 7b) has the most pronounced influence. Under lower congestion (85% of the baseline), carbon mitigation decreases by 11.9% in 2025 but decreases to a smaller deviation (−6.4%) by 2035, whereas higher congestion (115% of the baseline) increases mitigation by 12.8% in 2025 and +6.9% in 2035. These results highlight that traffic conditions directly shape the extent of emission reduction: when congestion is more severe, the comparative advantage of EVs over conventional vehicles becomes greater because of the higher energy efficiency of electrified drivetrains under stop-and-go conditions. From a policy perspective, this underscores the dual importance of traffic management. On the one hand, congestion alleviation remains a goal for improving overall urban liveability. On the other hand, residual congestion could paradoxically amplify the mitigation contribution of EVs relative to that of gasoline vehicles, suggesting that the carbon benefits of electrification are context dependent.

The charging convenience factor (Fig. 7c) also has substantial effects, with deviations ranging from −8.4% to +7.7% across the projection horizon. This reflects the extent to which charging accessibility shapes user behavior and the operational carbon profile of EVs. Greater charging convenience encourages higher electric mileage shares in PHEVs and EREVs, thereby enhancing carbon reduction. In Shanghai, where the density of charging facilities has expanded rapidly in recent years, these findings suggest that continued infrastructure deployment is essential for sustaining the city's mitigation trajectory. Moreover, the results align with empirical studies showing that users' charging frequency and reliance on electricity are strongly determined by the perceived accessibility of charging stations.

The power grid emission factor (Fig. 7d) represents another key driver. Adjustments of ±15% in grid carbon intensity lead to proportional shifts in mitigation performance, with impacts close to ±10%. As Shanghai's electricity mix remains partially dependent on coal, grid decarbonization is a decisive condition for unlocking the full potential of EVs. National energy transition plans envision a rapid scale-up of renewables and nuclear power by 2030, and the sensitivity results confirm that accelerated implementation of these measures could yield direct benefits for the transport sector. In particular, the alignment of transport electrification with grid decarbonization policies is critical for maximizing synergies across sectors.

Similarly, the electric intensity of EVs (Fig. 7e) has a significant influence, with deviations of approximately ±10%. A higher electric intensity translates into greater electricity consumption per vehicle kilometer, which reduces the relative mitigation advantage when the grid remains carbon intensive. Conversely, improvements in battery efficiency, vehicle lightweighting, and regenerative braking technologies can mitigate this effect. In the context of Shanghai, where the uptake of long-range BEVs with larger batteries has accelerated, policies promoting efficiency standards and technology innovation are essential to avoid rebound effects from increased energy demand.

Taken together, the sensitivity analysis demonstrates that while economic and behavioral factors such as fuel cost or travel demand have only minor effects, structural determinants linked to traffic conditions, charging infrastructure, and the carbon profile of the electricity system are decisive. These findings reinforce the notion that transport decarbonization cannot be addressed in isolation; instead, it must be embedded within a broader urban governance framework that simultaneously addresses congestion management, infrastructure planning, and energy system transformation. Accordingly, policy efforts in cities should prioritize congestion mitigation strategies, accelerate the deployment of user-friendly charging facilities, and ensure the alignment of EV promotion with the decarbonization of the regional power grid. By doing so, the city can maximize the long-term mitigation contribution of EVs and position itself as a leading example of integrated low-carbon urban transport development.

### *4.4. Policy implications*

The findings of this study have several policy implications for enhancing the decarbonization potential of EVs in Shanghai. First, the scenario analysis demonstrates that BEVs possess the greatest absolute mitigation capacity; however, their effectiveness is highly dependent on sustained policy support. Incentives such as purchase subsidies [57], license plate privileges, and charging infrastructure investment [58] remain crucial to secure continuous market expansion and prevent stagnation in the medium term.

In comparison, PHEVs have more modest benefits because of their continued reliance on gasoline. Nonetheless, their performance can be strengthened through targeted measures that encourage higher electric mileage shares [59], for instance, by tightening regulatory requirements or offering differentiated charging incentives. EREVs, in contrast, reveal a dynamic growth trajectory and may serve as an important transitional solution, especially in contexts where charging networks are unevenly distributed. Policy support for EREVs should therefore be tailored to specific niches, such as long-distance commuting or peri-urban areas [60, 61], to complement the large-scale adoption of BEVs without reinforcing long-term fossil fuel dependence.

Furthermore, the sensitivity analysis underscores that structural conditions, rather than purely economic drivers, are decisive for shaping long-term mitigation outcomes. Congestion, charging accessibility, and the carbon intensity of electricity emerge as the most influential variables [62, 63], suggesting that vehicle electrification must be embedded in a broader urban governance and energy transition framework [64]. This entails simultaneously promoting congestion management strategies, accelerating the deployment of accessible charging infrastructure [65], and aligning transport electrification with power system decarbonization [66].

Taken together, these results indicate that the carbon neutrality pathway should prioritize three interlinked dimensions: scaling up BEV adoption with robust policy backing, positioning EREVs and PHEVs as transitional yet complementary options, and embedding electrification policies within wider structural reforms in mobility and energy systems. This integrated strategy will enable cities not only to maximize the carbon reduction contribution of EVs but also to establish a replicable model for other global megacities seeking sustainable transport transitions.

## 5. Conclusion

In this study, a comprehensive assessment of the real-world energy consumption and operational carbon emissions of EVs in Shanghai from 2022-2024 was conducted. To achieve this goal, a bottom-up model incorporating vehicle sales, real-world energy intensity, and AVKT was developed to generate accurate and robust estimates of EV energy use and emissions. In parallel, an outlook model based on a system dynamics approach was established to project future decarbonization trajectories between 2025 and 2035 under four distinct policy scenarios. The key findings derived from this analysis are summarized as follows.

### *5.1. Core findings*

- **The real-world energy consumption of EVs in Shanghai was consistently higher than that of the official test cycle data, with BEVs showing an average discrepancy of 20.8%. For PHEVs and EREVs, this deviation was even more significant, with some models consuming up to 138% more energy than reported.** These findings highlight a critical gap between official efficiency standards and actual performance, particularly for PHEVs and EREVs, where official data often overestimate energy efficiency. In addition, vehicle type and weight impact energy use and carbon emissions, as compared with their sedan counterparts, SUVs tend to require at least 14% more energy. Additionally, a strong correlation was observed between vehicle weight and energy consumption, with heavier models—especially PHEVs and EREVs—exhibiting more variability in energy use owing to additional features and larger battery sizes.
- **BEVs dominate the Shanghai market and contribute the most absolute mitigation, with operational energy use increasing from 737 TJ in 2022 to 5999 TJ in 2024, driving associated emissions up to 975 kt$CO_2$ in 2024.** However, their climate benefits remain highly dependent on the carbon intensity of the electricity grid. Additionally, EREVs demonstrate a steadily expanding mitigation trajectory, supported by higher average utility factors (0.77–0.99) that indicate greater reliance on electricity. In contrast, PHEVs could only deliver comparatively modest reductions, playing a complementary role. Under the CP scenario, EREVs demonstrate the most significant decarbonization potential, reaching a carbon mitigation of 870 kt$CO_2$ in 2035, followed by PHEVs, with a more conservative improvement, contributing to a lower reduction in carbon emissions. In contrast, BEVs, which are approaching market saturation, are expected to experience fluctuating decarbonization performance. Despite BEVs being a dominant force in the reduction of emissions, their potential for further decarbonization will diminish as the market becomes saturated and their growth slows.

- **In terms of external factors, the congestion index and charging convenience have the greatest and second-largest effects (8.8% and 7.9%, respectively, on average), respectively, significantly enhancing the carbon reduction potential.** In addition, the power grid emission factor and electricity also have direct and observable effects on future decarbonization. These findings highlight the importance of integrated policies addressing urban traffic, charging infrastructure, and energy systems to maximize the carbon reduction potential of EVs.

### *5.2. Future work*

Future research should expand the geographical scope beyond Shanghai to capture interregional variations in grid carbon intensity, traffic conditions, and consumer behavior. The incorporation of life-cycle emissions from manufacturing, recycling, and end-of-life stages will also provide a more holistic understanding of EV decarbonization potential. In addition, integrating real-time big data on driving patterns and charging behavior could refine the accuracy of operational emission estimates. Finally, scenario modeling that explicitly couples transportation electrification with broader energy system transitions is critical for guiding policy design. Such efforts will help identify optimal pathways for scaling up EVs while ensuring that their deployment aligns with long-term carbon neutrality goals.

## Acknowledgments

None.

## References

[1] IEA (2025), Global EV Outlook 2025, IEA, Paris https://www.iea.org/reports/global-ev-outlook-2025, License: CC BY 4.0

[2] Wang H, Ou X, Zhang X. Mode, technology, energy consumption, and resulting CO2 emissions in China's transport sector up to 2050. Energy Policy 2017;109:719-733.

[3] IEA (2024), Global EV Outlook 2024, IEA, Paris https://www.iea.org/reports/global-ev-outlook-2024, License: CC BY 4.0

[4] Chen W, Sun X, Liu L, Liu X, Zhang R, Zhang S, et al. Carbon neutrality of China’s passenger car sector requires coordinated short-term behavioral changes and long-term technological solutions. One Earth 2022;5:875-891.

[5] Gan Y, Lu Z, He X, Hao C, Wang Y, Cai H, et al. Provincial Greenhouse Gas Emissions of Gasoline and Plug-in Electric Vehicles in China: Comparison from the Consumption-Based Electricity Perspective. Environmental Science & Technology 2021;55:6944-6956.

[6] Ren Y, Sun X, Wolfram P, Zhao S, Tang X, Kang Y, et al. Hidden delays of climate mitigation benefits in the race for electric vehicle deployment. Nature Communications 2023;14:3164.

[7] Wang Y, Ma M, Zhou N, Ma Z. Paving the way to carbon neutrality: Evaluating the decarbonization of residential building electrification worldwide. Sustainable Cities and Society 2025;130:106549.

[8] Yuan H, Ma M, Zhou N, Xie H, Ma Z, Xiang X, et al. Battery electric vehicle charging in China: Energy demand and emissions trends in the 2020s. Applied Energy 2024;365:123153.

[9] Song J, Cha J, Choi M. A study on 5-cycle fuel economy prediction model of electric vehicles using numerical simulation. Energy 2024;309:133189.

[10] Abdelaty H, Al-Obaidi A, Mohamed M, Farag HEZ. Machine learning prediction models for battery-electric bus energy consumption in transit. Transportation Research Part D: Transport and Environment 2021;96:102868.

[11] Zhu Q, Huang Y, Lee CF, Liu P, Zhang J, Wik T. Predicting Electric Vehicle Energy Consumption From Field Data Using Machine Learning. IEEE Transactions on Transportation Electrification 2025;11:2120-2132.

[12] Qiao Q, Zhao F, Liu Z, He X, Hao H. Life cycle greenhouse gas emissions of Electric Vehicles in China: Combining the vehicle cycle and fuel cycle. Energy 2019;177:222-233.

[13] Deng Y, Ma M, Zhou N, Ma Z, Yan R, Ma X. China's plug-in hybrid electric vehicle transition: An operational carbon perspective. Energy Conversion and Management 2024;320:119011.
[14] Dong Y, Guo X, Wang M, Xu J. Modeling carbon intensity of electric vehicles in the well-to-wheels phase under different traffic flow conditions. iScience 2024;27:111070.
[15] Xu X, Aziz HMA, Guensler R. A modal-based approach for estimating electric vehicle energy consumption in transportation networks. Transportation Research Part D: Transport and Environment 2019;75:249-264.
[16] Liu X, Elgowainy A, Vijayagopal R, Wang M. Well-to-Wheels Analysis of Zero-Emission Plug-In Battery Electric Vehicle Technology for Medium- and Heavy-Duty Trucks. Environmental Science & Technology 2021;55:538-546.
[17] Zhang C, Yang F, Ke X, Liu Z, Yuan C. Predictive modeling of energy consumption and greenhouse gas emissions from autonomous electric vehicle operations. Applied Energy 2019;254:113597.
[18] Sun D, Zheng Y, Duan R. Energy consumption simulation and economic benefit analysis for urban electric commercial-vehicles. Transportation Research Part D: Transport and Environment 2021;101:103083.
[19] Basso R, Kulcsár B, Egardt B, Lindroth P, Sanchez-Diaz I. Energy consumption estimation integrated into the Electric Vehicle Routing Problem. Transportation Research Part D: Transport and Environment 2019;69:141-167.
[20] Zhang R, Yao E. Mesoscopic model framework for estimating electric vehicles' energy consumption. Sustainable Cities and Society 2019;47:101478.
[21] García-Afonso Ó. Impact of powertrain electrification on the overall CO2 emissions of intercity public bus transport: Tenerife Island test case. Journal of Cleaner Production 2023;412:137365.
[22] Yan R, Zhou N, Ma M, Mao C. India's residential space cooling transition: Decarbonization ambitions since the turn of millennium. Applied Energy 2025;391:125929.
[23] Wu Z, Wang C, Wolfram P, Zhang Y, Sun X, Hertwich E. Assessing electric vehicle policy with region-specific carbon footprints. Applied Energy 2019;256:113923.
[24] Yang L, Yu B, Yang B, Chen H, Malima G, Wei Y-M. Life cycle environmental assessment of electric and internal combustion engine vehicles in China. Journal of Cleaner Production 2021;285:124899.
[25] Petrauskienė K, Skvarnavičiūtė M, Dvarionienė J. Comparative environmental life cycle assessment of electric and conventional vehicles in Lithuania. Journal of Cleaner Production 2020;246:119042.

[26] Shafique M, Luo X. Environmental life cycle assessment of battery electric vehicles from the current and future energy mix perspective. Journal of Environmental Management 2022;303:114050.
[27] Li F, Ou R, Xiao X, Zhou K, Xie W, Ma D, et al. Regional comparison of electric vehicle adoption and emission reduction effects in China. Resources, Conservation and Recycling 2019;149:714-726.
[28] Zhang H, Xue B, Li S, Yu Y, Li X, Chang Z, et al. Life cycle environmental impact assessment for battery-powered electric vehicles at the global and regional levels. Scientific Reports 2023;13:7952.
[29] Ma M, Zhou N, Feng W, Yan J. Challenges and opportunities in the global net-zero building sector. Cell Reports Sustainability 2024;1:100154.
[30] Jiang Y, Guo J, Zhao D, Li Y. Intelligent energy consumption prediction for battery electric vehicles: A hybrid approach integrating driving behavior and environmental factors. Energy 2024;308:132774.
[31] Zhang S, Ma M, Zhou N, Yan J, Feng W, Yan R, et al. Estimation of global building stocks by 2070: Unlocking renovation potential. Nexus 2024;1:100019.
[32] Jauhar SK, Sethi S, Kamble SS, Mathew S, Belhadi A. Artificial intelligence and machine learning-based decision support system for forecasting electric vehicles' power requirement. Technological Forecasting and Social Change 2024;204:123396.
[33] Basso R, Kulcsár B, Sanchez-Diaz I. Electric vehicle routing problem with machine learning for energy prediction. Transportation Research Part B: Methodological 2021;145:24-55.
[34] Jia Z, Yin J, Cao Z, Wu L, Wei N, Zhang Y, et al. Regional vehicle energy consumption evaluation framework to quantify the benefits of vehicle electrification in plateau city: A case study of Xining, China. Applied Energy 2025;377:124626.
[35] Xu B, Sharif A, Shahbaz M, Dong K. Have electric vehicles effectively addressed CO2 emissions? Analysis of eight leading countries using quantile-on-quantile regression approach. Sustainable Production and Consumption 2021;27:1205-1214.
[36] Xue M, Lin B-L, Tsunemi K. Emission implications of electric vehicles in Japan considering energy structure transition and penetration uncertainty. Journal of Cleaner Production 2021;280:124402.
[37] Hofmann J, Guan D, Chalvatzis K, Huo H. Assessment of electrical vehicles as a successful driver for reducing CO2 emissions in China. Applied Energy 2016;184:995-1003.
[38] Lu Q, Duan H, Shi H, Peng B, Liu Y, Wu T, et al. Decarbonization scenarios and carbon reduction potential for China’s road transportation by 2060. npj Urban Sustainability 2022;2:34.

[39] Sun D, Kyere F, Sampene AK, Asante D, Kumah NYG. An investigation on the role of electric vehicles in alleviating environmental pollution: evidence from five leading economies. Environmental Science and Pollution Research 2023;30:18244-18259.

[40] Ma M, Zhang S, Liu J, Yan R, Cai W, Zhou N, et al. Building floorspace and stock measurement: A review of global efforts, knowledge gaps, and research priorities. Nexus 2025;2:100075.

[41] Al-Wreikat Y, Serrano C, Sodré JR. Effects of ambient temperature and trip characteristics on the energy consumption of an electric vehicle. Energy 2022;238:122028.

[42] Ghanbari Motlagh S, Oladigbolu J, Li L. A review on electric vehicle charging station operation considering market dynamics and grid interaction. Applied Energy 2025;392:126058.

[43] Shariatzadeh M, Lopes MAR, Henggeler Antunes C. Electric vehicle users' charging behavior: A review of influential factors, methods and modeling approaches. Applied Energy 2025;396:126167.

[44] Barakat S, Guven AF, Abdelaziz AY, Samy MM. A comprehensive review of electric vehicles and sustainable urban mobility in the Middle East and North Africa. Renewable and Sustainable Energy Reviews 2026;225:116154.

[45] Gathen L, Upham P, Newig J. System capacities and implementation challenges in Germany's electric vehicle policy mix: an actor-centered review of the policy mix. Renewable and Sustainable Energy Reviews 2025;218:115787.

[46] da Costa VBF, Bitencourt L, Dias BH, Soares T, de Andrade JVB, Bonatto BD. Life cycle assessment comparison of electric and internal combustion vehicles: A review on the main challenges and opportunities. Renewable and Sustainable Energy Reviews 2025;208:114988.

[47] Choi M, Cha J, Song J. Impact of lightweighting and driving conditions on electric vehicle energy consumption: In-depth analysis using real-world testing and simulation. Energy 2025;323:135746.

[48] Huang H, He H, Wang Y, Zhang Z, Wang T. Energy consumption prediction of electric vehicles for data-scarce scenarios using pre-trained model. Transportation Research Part D: Transport and Environment 2025;146:104830.

[49] Lei N, Zhang H, Hu J, Hu Z, Wang Z. Sim-to-real design and development of reinforcement learning-based energy management strategies for fuel cell electric vehicles. Applied Energy 2025;393:126030.

[50] Li J, Yi Q, Zhu P, Hu J, Yi S. Data-driven co-optimization method of eco-adaptive cruise control for plug-in hybrid electric vehicles considering risky driving behaviors. Applied Energy 2025;392:126039.

[51] Sandhiya E, Gajanand MS. Hybrid charging infrastructure for electric vehicles: A comprehensive review and avenues for future research. Applied Energy 2025;401:126749.

[52] Huang H, Gao K, Wang Y, Najafi A, Zhang Z, He H. Sequence-aware energy consumption prediction for electric vehicles using pre-trip realistically accessible data. Applied Energy 2025;401:126673.

[53] Gao J, Xu X, Zhang T. Forecasting the development of Clean energy vehicles in large Cities: A system dynamics perspective. Transportation Research Part A: Policy and Practice 2024;181:103969.

[54] Liu D, Xiao B. Exploring the development of electric vehicles under policy incentives: A scenario-based system dynamics model. Energy Policy 2018;120:8-23.

[55] Feng B, Ye Q, Collins BJ. A dynamic model of electric vehicle adoption: The role of social commerce in new transportation. Information & Management 2019;56:196-212.

[56] Raoofi Z, Mahmoudi M, Pernestål A. Electric truck adoption and charging development: Policy insights from a dynamic model. Transportation Research Part D: Transport and Environment 2025;139:104515.

[57] Bauer J, Letmathe P, Woeste R. Total cost of ownership for battery electric vehicles: The role of energy prices. Applied Energy 2025;389:125764.

[58] Javazi L, Alinaghian M, Khosroshahi H. Evaluating government policies promoting electric vehicles, considering battery technology, energy saving, and charging infrastructure development: A game theoretic approach. Applied Energy 2025;390:125799.

[59] Li D, Hu Q, Jiang W, Dong H, Song Z. Integrated power and thermal management for enhancing energy efficiency and battery life in connected and automated electric vehicles. Applied Energy 2025;396:126213.

[60] Sharma RB, Majumdar BB, Maitra B. Commuter and non-commuter preferences for plug-in hybrid electric vehicle: A case study of Delhi and Kolkata, India. Research in Transportation Economics 2024;103:101415.

[61] Shu C, Bao Y, Gao Z, Gao Z. Exploring Electric Vehicle Purchases and Residential Choices in a Two-Dimensional Monocentric City: An Agent-Based Microeconomic Model. Engineering 2025;46:316-330.

[62] Du B, Jia H, Zhang B, Li Y, Wang H, Yan J, et al. Evaluating the carbon-emission reduction potential of employing low-carbon demand response to guide electric-vehicle charging: A Chinese case study. Applied Energy 2025;397:126196.

[63] Woo S, Strobel L, Yuan Y, Pruckner M, Lipman TE. Exploring bidirectional charging strategies for an electric vehicle population. Applied Energy 2025;397:126361.

[64] Dik A, Sun H, Calautit JK, Kutlu C, Boukhanouf R, Omer S. Sustainable urban energy dynamics: Integrating renewable energy and electric vehicles in a European context. Energy Conversion and Management 2025;340:119972.

[65] Deveci M, Erdogan N, Pamucar D, Kucuksari S, Cali U. A rough Dombi Bonferroni based approach for public charging station type selection. Applied Energy 2023;345:121258.

[66] Huang P, Sandström M. Do smart charging and vehicle-to-grid strengthen or strain power grids with rising EV adoption? Insights from a Swedish residential network. Applied Energy 2025;401:126713.